\documentclass[prl,showpacs,twocolumn]{revtex4}
\usepackage{tabularx}
\usepackage{amsmath}
\usepackage{amssymb}
\usepackage{graphicx}
\usepackage{dcolumn}
\usepackage{bm}
\usepackage{ulem}
\usepackage{color}

\begin{document}
\title{Pairing Symmetry, Phase diagram and Edge Modes in Topological Fulde-Ferrell-Larkin-Ovchinnikov Phase}
\author{Chun Fai Chan}
\author{Ming Gong}
\email{skylark.gong@gmail.com}
\affiliation{Department of Physics and Centre for Quantum Coherence, The Chinese University of Hong Kong, Shatin, N.T., Hong Kong, China}

\begin{abstract}
The realizations of spin-orbit coupling in cold atoms lead to a burst of research activities in the searching of topological 
matters in ultracold atom systems. The very recent theoretical predictions show that topological Fulde-Ferrell-Larkin-Ovchinnikov (FFLO)
superfluids can be realized with proper spin-orbit coupling and Zeeman fields. In this work, a comprehensive understanding of the pairing symmetry, 
phase diagram and the edge modes in this new topological matter are presented.  The momentum of the Cooper pairs plays the role of 
renormalizing the in-plane Zeeman field and chemical potential. The in-plane Zeeman field and finite momentum pairing induce asymmetry to the effective 
$p$-wave pairing, apart from a small fraction of higher orbital components. The phase diagram is composed by different phases, which are determined by the 
topology and band gap nature of the superfluids. Especially, the gapped and gapless topological FFLO phase have totally different finite size effect. 
These novel features show that the spin-orbit coupled cold atoms provides an important platform in realizing topological matters which may not be
materialized with solids.
\end{abstract}

% Ming Gong verifed at 7.1.2013, please select several proper codes from the following candidates.
% 03.75.Ss Degenerate Fermi gases
% 74.20.Fg: BCS theory and its development
% 74.70.Tx: Heavy-fermion superconductors
% 74.25.Dw: Superconductivity phase diagrams
% 03.67.Lx Quantum computation architectures and implementations
% 03.65.Vf Phases: geometric; dynamic or topological
% 74.20.-z Theories and models of superconducting state

\maketitle

\begin{figure}
\centering
\includegraphics[width=3in]{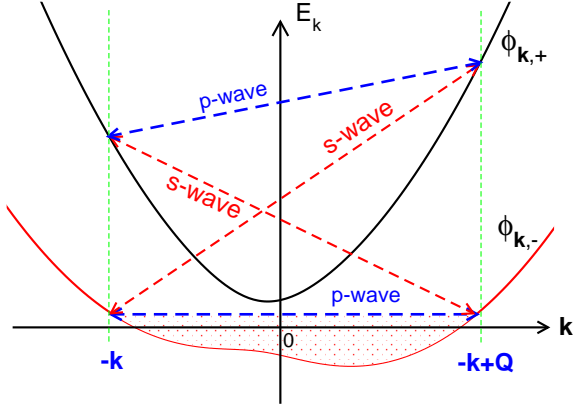}
\caption{(Color online). Single particle band structures with in-plane Zeeman field. The blue bidirectional arrows mark the possible
$p$-wave pairing in the same band, and the red bidirectional arrows mark the $s$-wave pairing between different bands. The two different 
bands have different chirality ($s = \pm$) or geometry phase\cite{Zhang,chunlei_natc13}.}
\label{fig1}
\end{figure}

The experimental realizations of spin-orbit coupling (SOC) in cold atom systems\cite{lin_nat11, pengjun_prl12,cheuk_prl12} lead 
to a burst of research activities in the searching of  novel new physics with this new interaction in the context of 
atomic, molecular and optical physics\cite{gong_prl11,han_pra12,seo_pra12,zengqiang_prl11,zhou_pra11,jiang_pra11, wu_prl13,zheng_pra13, Zheng_NJP, chen_pra12,
chunlei_natc13, Wei_natc13, Xiaji13, chun13, gong_prl12,renyuan_prl12,hu_njp13,lianyi_pra13,liu_prl12,seo_pra13,wei_pra13}, among which the 
topological matters are probably the most widely pursued phenomena. This is not strange because SOC is always the most unusual interaction in 
condensed matter physics with nontrivial topology or geometry phase, see Ref. [\onlinecite{R1, R2, R3}]. Now it is widely
recognized that the SOC and Zeeman field in degenerate Fermi gas can lead to topological Bardeen-Cooper-Schrieffer (BCS) phase when
$h^2 > \mu^2 + \Delta^2$,
where $h$, $\mu$, $\Delta$ are the Zeeman field strength (independent of its direction), chemical potential and order parameter, respectively. 
The basic reason is that SOC and Zeeman field can lead to equivalent $k_x + ik_y$ triplet pairing in the dressed basis
(eigenvectors of the single particle Hamiltonian, see Fig. \ref{fig1}) under proper condition. This idea, first unveiled by
Zhang\cite{Zhang} and Sato\cite{Sato} in cold atoms, can be tracked back to the work by Gor'kov and Rashba\cite{Rashba}, 
who showed that, due to inversion symmetry breaking induced SOC, the singlet and triplet pairing are mixed in the wave function of the Cooper
pairs in non-centrosymmetric superconductors. It is then applied to solid materials to search Majorana
Fermion, see the recent review and references therein\cite{Alicea}. This topological boundary put a strong constraint to the
realization of Majorana Fermions in practical experiments because the Cooper pair may be destroyed by strong 
Zeeman field due to Pauli depairing effect. Moreover, in cold atom systems, the topological phase is only possible to be observed in the strong 
coupling regime\cite{gong_prl11}. So the realization of topological phase  is always accompanied by  competition with either Pauli depairing effect or strong coupling effect. 

Generally speaking, there are two totally different paradigms to destroy the Cooper pair by strong Zeeman field. The most conventional way, for
example, in type-I superconductors, is the direct destroying of Cooper pair via a second-order phase transition. The other paradigm is that the strong 
Zeeman field first induce spontaneous translation symmetry breaking to the uniform superconductor and then destroy it via a first order phase transition. 
The latter case may be encountered in some type-II superconductors where magnetic and superconducting order coexist, which is now
widely known as the Fulde-Ferrell-Larkin-Ovchinnikov (FFLO) phase. Recent progresses show that the SOC and in-plane Zeeman field can provide an more
efficient route\cite{zheng_pra13, Zheng_NJP} to realize FFLO superfluids in ultracold atoms. The basic idea is quite simple, the in-plane
Zeeman field can deform the Fermi surface, making the single particle bands lack inversion symmetry. As a result, it is impossible
to find two fermions with opposite momentum but the same energy in the same band. Mathematically, we can write down the thermaldynamical potential 
as $\Omega({\bf Q}) = \Omega(0) + \nabla \Omega(0)\cdot {\bf Q} + \mathcal{O}({\bf Q}^2)$ in the small ${\bf Q}$ limit, and we can proof exactly
that $\nabla \Omega(0)\ne 0$ in the present of both in-plane Zeeman field and SOC\cite{MGnote}. The
small ${\bf Q}$ will not change the topology of the Fermi surface, thus in some parameter regime the topological FFLO
can be realized, as predicted in recent works\cite{chunlei_natc13,Wei_natc13,Xiaji13, chun13}. From the language of topology, this new
phase is just a deformed BCS phase.  In this new phase, all the physical parameters
are involved to determine the topological boundaries. Thus in principle, such a topological phase is a more suitable candidate for engineering topological phase transition in ultracold atoms. 

The physics of this new phase is still not well understood.  In this work, we aim to provide a comprehensive understanding of this new phase recently predicted 
in Ref. [\onlinecite{chunlei_natc13,Wei_natc13,Xiaji13, chun13}].  
Our major observations are summarized as following: (I) The finite momentum of the Cooper pairs plays the role of renormalizing the 
in-plane Zeeman field and chemical potential. (II) The effective pairing in the topological FFLO phase regime is $p$-type plus a
small fraction of higher orbital components. The effective in-plane Zeeman fields introduce asymmetry to the $p$-wave pairing, which can influence the properties
of the superfluids. However, this asmmetry will not change the Fermi surface topology, which is only determined by the strength of the effective Zeeman field. 
(III) The phase diagram composed by different phases is determined by topology and band gap nature (gapped or gapless). The gapped and gapless topological phase
have totally different finite size effect. These novel features in ultracold atom systems have no counterpart in solid materials.

\begin{figure}
\centering
\includegraphics[width=3in]{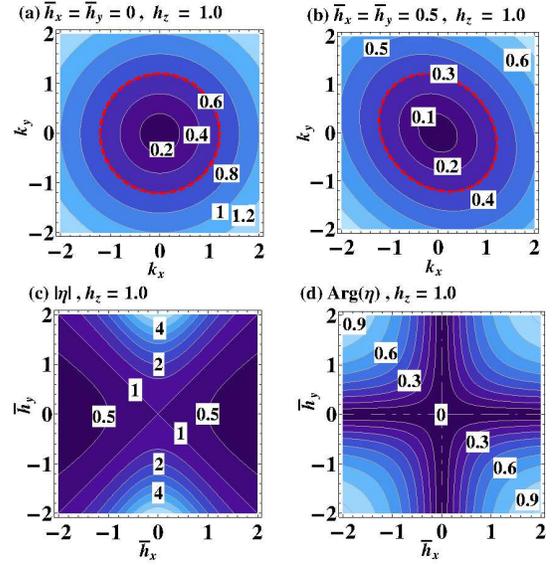}
\caption{(Color online). (a)-(b) the pairing function $|A_{{\bf k},{\bf Q}}|$ in momentum space. (c)-(d) the absolute value and phase (Arg) of $\eta$ as a function
of in-plane Zeeman field. }
\label{fig2}
\end{figure}

In the past two years great theoretical endeavors have been made trying to understand the physics in spin-orbital degenerate Fermi
gas based on self-consistent calculations\cite{gong_prl11,han_pra12,seo_pra12,zengqiang_prl11,zhou_pra11,jiang_pra11, wu_prl13,zheng_pra13, Zheng_NJP, chen_pra12, chunlei_natc13, Wei_natc13, Xiaji13, chun13, gong_prl12,renyuan_prl12,hu_njp13,lianyi_pra13,liu_prl12,seo_pra13,wei_pra13}. The triplet pairing induced by SOC is essential to overcome the Chandrasekhar-Clogston limit\cite{CC1, CC2}, thus
finite pairing can survive in the topological regime ($h^2 > \Delta^2 + \mu^2$). The essential physics, however, can be well understood without these
complicated calculations, therefore in the coming discussion, we mainly consider the following
Bogliubov-de Gennes (BdG) Hamiltonian\cite{zheng_pra13, Zheng_NJP, chunlei_natc13},
\begin{equation}
H_{\text{BdG}}(\mathbf{k})=%
\begin{pmatrix}
H_{0}({\frac{\mathbf{Q}}{2}}+\mathbf{k}) & \Delta \\
\Delta & -\sigma _{y}H_{0}^{\ast }({\frac{\mathbf{Q}}{2}}-\mathbf{k})\sigma
_{y}%
\end{pmatrix}%
,  \label{eq-bdg}
\end{equation}%
where the Nambu basis is chosen as $(c_{\mathbf{k}+\mathbf{Q}/2,\uparrow}, c_{\mathbf{k}+\mathbf{Q}/2,\downarrow }, 
c_{-\mathbf{k}+\mathbf{Q}/2,\downarrow }^{\dagger }, -c_{-\mathbf{k}+\mathbf{Q}/2,\uparrow }^{\dagger})^{T}$, with 
$c_{{\bf k}s}$ the destructive operator and ${\bf Q}=(Q_x, Q_y)$ the total Cooper pair momentum. $H_0$ is the single particle Hamiltonian,
\begin{equation}
H_0 = {\bf k}^2 + \alpha (k_x\sigma_y - k_y\sigma_x) + h_z\sigma_z + h_x\sigma_x + h_y \sigma_y,
\end{equation}
where ${\bf k} = (k_x, k_y)$, $\alpha$ is the Rashba SOC coefficient, and $h_x$, $h_y$, $h_z$ are Zeeman field strength, and $\sigma_x$, $\sigma_y$ and $\sigma_z$ are 
Pauli matrices.  Notice that the energy is scaled by Fermi energy $E_{\text{F}}$, and the momentum is scaled by the corresponding Fermi momentum $k_{\text{F}}$.  This model can be 
readily realized with degenerate Fermi gas with Rashba SOC. Notice that while the physics in solid materials and cold atoms are essentially identical, 
their physical parameters are chosen in totally different regions. In this work we mainly discuss the physics in cold atoms, thus all the energy parameters 
are comparable in magnitude\cite{Solids}. 

\begin{figure}
      \centering
      \includegraphics[width=3in]{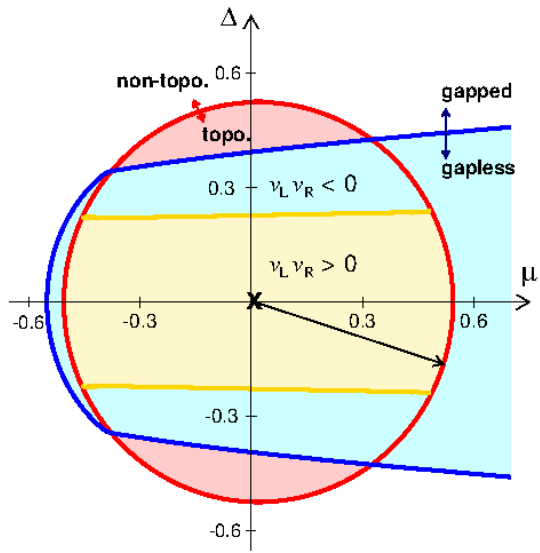}
      \caption{(Color online). Phase diagram in the $\mu-\Delta$ plane. The solid circle represents the boundary between topological phase and trivial phase, 
      and the open parabola represents the boundary between gapless phase and gapped phase. The overlaps between these two curves define different phases. In the
      topological gapless phase regime, the (almost) horizontal line define two different phases with 
      counter-propagating ($v_{\text{L}}v_{\text{R}} < 0$) edge modes and  co-propgatating 
      ($v_{\text{L}}v_{\text{R}} > 0$) edge modes. This phase diagram is obtained using parameters $\alpha = 0.5$, $h_x =0.2$, $h_y = 0.1$, $h_z = 0.5$, $Q_x = 0.3$ , $Q_y = 0.2$.}
      \label{fig3}
\end{figure}

{\it Pairing symmetry.-} Firstly, the topology of the Fermi surface is determined by solely the physics at zero momentum.  In this case the influence of the 
Cooper pair momentum ${\bf Q}$ can be fully absorbed by defining two effective in-plane Zeeman fields $\bar{h}_x = h_x - \alpha Q_y/2$, 
$\bar{h}_y = h_y + \alpha Q_x/2$ and an effective chemical potential $\bar{\mu} = \mu - E_Q$, with $E_Q = (Q_x^2 +Q_y^2)/4$ the corresponding kinetic 
energy of the Cooper pair.  For convenience, we also 
define Zeeman field strength and effective Zeeman field strength as $h^2 = h_x^2 + h_y^2 + h_z^2$ and $\bar{h}^2 = h_z^2 + \bar{h}_x^2 + \bar{h}_y^2$, respectively. 
The eigenvectors of the single particle Hamiltonian $H_0 \phi_{{\bf k}s} = \xi_{{\bf k}s}\phi_{{\bf k}s}$, where $s = \pm$ define the chirality of the bands. 
$\phi_{{\bf k}s}$, hereafter, is called as dressed basis. Assume $\phi_{{\bf k},s} = (a_{{\bf k}s} c_{{\bf k},\uparrow} + c_{{\bf k},\downarrow})/\sqrt{(|a_{{\bf k}s}|^2 + 1}$, with 
$a_{{\bf k}s} = (s\bar{h} - h_z)/(\bar{h}_x + i\bar{h}_y)$. 
To observe the topological phase, the chemical potential should cut only  one band, see Fig. 1. In the dressed basis, we find 
$\langle \phi_{{\bf k}+{\frac{{\bf Q}}{2}},-}\phi_{{\bf -k}+ {{\bf Q} \over 2},-} \rangle = A_{{\bf k}, {\bf Q}} \Delta$,
where $\Delta$ is the $s$-wave pairing strength and the pairing function reads as
\begin{equation}
A_{{\bf k}, {\bf Q}} = {a_{{\bf k+Q}/2-} - a_{{\bf -k+Q}/2-} \over \sqrt{(|a_{{\bf k+Q}/2-}|^2 + 1)(|a_{{\bf -k+Q}/2-}|^2 + 1}}.
\end{equation}
Notice that the numerator of $A_{{\bf k}, {\bf Q}}$ is an odd function with respect to ${\bf k}$, while the denominator is an even function, thus we have
\begin{equation}
A_{{\bf k}, {\bf Q}} = \alpha (ik_x + \eta k_y) f(k^2) + \cdots,
\label{eq-effp}
\end{equation}
where all the even order terms ($s$-wave and $d$-wave term) are absent, and the suspension point represent the contribution of higher 
orbital components. Notice that $f(k^2)$, which may be a complex number here, is a function of $k^2$, thus a small fraction  of higher orbital components have been absorbed 
into the first term. Eq. \ref{eq-effp} is the major observation in this work. We can recover the well-known $k_x + ik_y$ 
pairing at the limit without in-plane Zeeman field, in which condition $f(k^2) = 1/\sqrt{h_z^2 + \alpha^2k^2}$. 
Obviously, $\eta$ controls the asymmetry of the superfluids.  Similar analysis show
that all the intra-band couplings should be an odd function thus $p$-wave type, while the inter-band pairing can be $s$-wave type in
the small momentum limit, see schematically shown in Fig. \ref{fig1}. The pairing symmetry for other types of SOC and Zeeman field can also be examined
in a simialr way. 

We have verified that the first term is always dominant over the second term for the typical parameter regimes in cold atoms. In actually, we can easily
check that when $\alpha k \gg \bar{h}, h$, $A_{{\bf k}, {\bf Q}} \sim  (k_x + ik_y)/|{\bf k}|$. In this sense, we can conclude that the effective in-plane
Zeeman field generally play the role of inducing asymmetry to the effective pairing, which is controlled by $\eta$ in Eq. \ref{eq-effp}. In Fig. \ref{fig2}a - b we plot the pairing function in momentum space, which show clearly asymmetry due to the presents of in-plane Zeeman field. In Fig. \ref{fig2}c-d, we plot the asymmetry
factor $\eta$ as a function of in-plane Zeeman field. We see that $|\eta| = 1$ only when $|\bar{h}_x| = |\bar{h}_y|$, however, at this condition, its
phase still depends strongly on the in-plane Zeeman field. When $|\bar{h}_y| \gg |\bar{h}_x|$, we find $|\eta| \ll 1$, and the superfluids is continuously
tuned from $k_x + ik_y$ superfluids to $k_y$ superfluids. Similarly, the $k_x$ superfluids can be obtained in the limit when $|\bar{h}_y| \ll |\bar{h}_x|$.  Here we need to 
emphasize that the asymmetry of the superfluids just changes the properties of the superfluids (gapped or gapless, {\it etc});
however, it will not change its topology, which is determined solely by the effective Zeeman field strength, see Eq. \ref{eq-topo2}.

\begin{figure}
      \centering
      \includegraphics[width=3in]{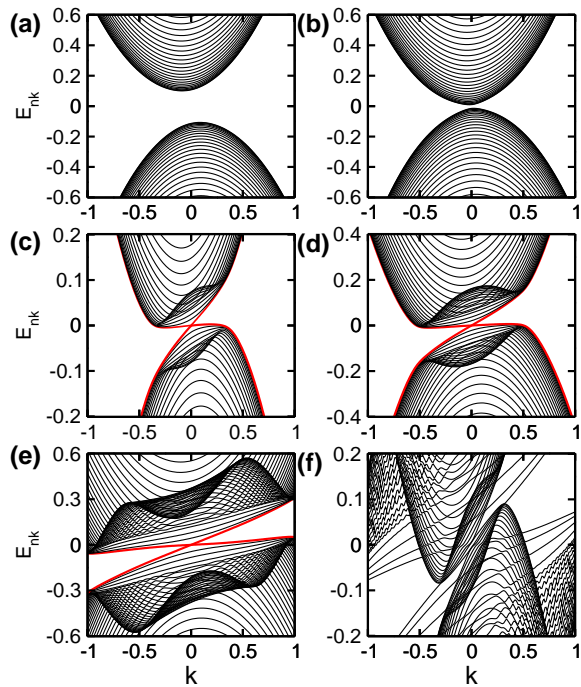}
      \caption{(Color online). Typical band structure in a stripe geometry with width $L = 100$. From (a) - (f), $\mu$ = -0.6, -0.5, -0.4, -0.3, 0.3, and 0.7, 
      respectively. $\Delta = 0.2$, and for vaues of other parameters see Fig. \ref{fig3}. In (c) - (f), the red solid lines represent the edge modes.}
\label{fig4}
\end{figure}

{\it Phase diagram.-} The out-of-plane Zeeman field and SOC is essential to realize a single band with non-trivial
Berry phase, which is prerequisite for the realization of topological matters with $s$-wave interactions. The topological phase transition can be determined the 
Pfaffian of the Hamiltonian. We can define a topological index $\mathcal{M} = \text{sign}$(Pf$(\Gamma (0))$), where $\Gamma(\mathbf{k}) = -i H_{\text{BdG}}(\mathbf{k})\Lambda$, and
$\Lambda$ is the standard particle-hole symmetry operator without the complex conjugate operator\cite{ghost_prb10}. $\mathcal{M} = -1$ corresponds to topological phase.
We determine the topological boundary as 
\begin{equation}
\bar{\mu}^2  + \Delta^2 < h_z^2 + \bar{h}_x^2 + \bar{h}_y^2.
\label{eq-topo2}
\end{equation}
The above result is further verified by Chern number for all the filled bands $\mathcal{C} = \sum_n \mathcal{C}_n$, where $\mathcal{C}_n = -\text{Im}{1\over 2\pi}\int d^2k \langle \partial_{k_x}\psi_{n{\bf k}}|\partial_{k_y}\psi_{n{\bf k}} \rangle$ and $|\psi_{n{\bf k}}\rangle$ the eigenvalues of the BdG equation. We find $\mathcal{C} = 1$ 
in the topological regime defined above, and $\mathcal{C} = 0$ otherwise.
We see that the topological boundary is identical to that for topological BCS superfluids with out-of-plane Zeeman field\cite{Alicea, gong_prl11}, 
except that now the Zeeman field should be replaced by the effective Zeeman field.  This topological boundary condition is plotted in Fig. \ref{fig3} by a perfect circle with
radius $R = \bar{h}$ (notice that the center of the circle is shifted by the kinetic energy of Cooper pair $E_Q$). However due to the present of  ${\bf k}\cdot {\bf Q}$ term in the diagonal 
term in Eq. \ref{eq-bdg}, the system can become gapless at ${\bf k} \ne 0$. This is a general feature of the FFLO phase both in solids
and cold atoms.  The analytical eigenvalues for the original Hamiltonian can not be obtained anymore. The boundary for gappless and gapped phase 
(defined by Det($H_{\text{BdG}}({\bf k\ne 0})$) =0) is determined numerically in Fig. \ref{fig3}, which is represented by a right open parabola. The different phases
in the parameter space thus can be defined by the overlaps between these two curves. Notice that both the topological phase
and trivial phase can be either gapped or gapless. Here the gapped FFLO phase is defined as a separate phase because in this regime, the Majorana Fermion
can be realized at the vortex core, which is protected by a large fundamental gap. The gapless FFLO phase is not suitable for this particular application, however, it
possesses some intriguing features which are beyond the accessibility of its counterpart in solid materials, see below. 

{\it Edge modes and finite size effect-}  
It is well-known in condensed matter community that topological gapless edge modes may emergent at the boundary of topological matters due to
the bulk-edge correspondence\cite{BE}. In other words, the edge modes are a manifestation of bulk topology.
Such topological protected edge modes have been widely discussed in the context of topological insulators\cite{Edge1, Edge2, Edge3} and quantum spin Hall 
effect\cite{S1,S2,S3,S4}, 
however, direct
observation of the topologically protected edge modes are always a great challenge in solid materials. In this sense, it is important to study the topologically protected edge modes in topological BCS and topological FFLO superconductors in ultracold atom systems, which can be directly probed in cold atom systems using momentum resolved radio frequency spectroscopy\cite{rf1, rf2, rf3}.

\begin{figure}
      \centering
      \includegraphics[width=3in]{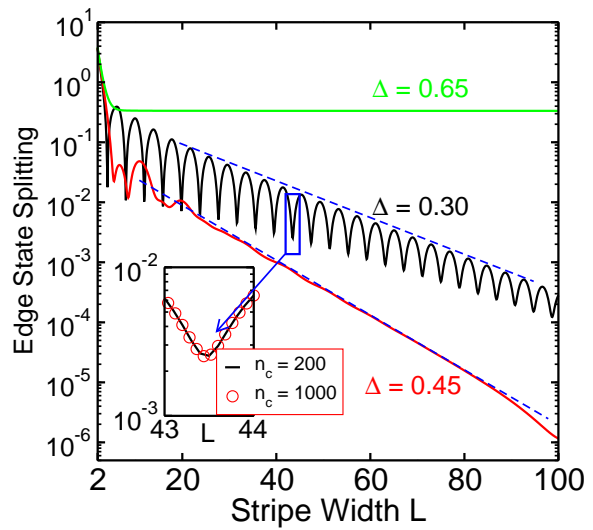}
      \caption{(Color online). Finite size effect for localized edge modes. $\mu = 0.1$ and for the values of other parameters, see Fig. \ref{fig3}. The gap width 
      for trivial gapped phase ($\Delta = 0.65$) is also presented for comparison.}
\label{fig5}
\end{figure}

We present the major results for a stripe geometry along $x$ direction in Fig. \ref{fig4} for fixed chemical potential. The stripe width along the perpendicular direction is set to $L = 100$, which is large enough to ensure that the coupling between the two edges are vanishing small.  Hard wall boundary condition is used. We numerically solve the BdG equation using plane wave basis by assuming $\Psi_k = e^{ikx} \sum_n c_n \phi_n(y)$,  where $\phi_n(y) = \sin(n\pi y/L)$, with $n$ smaller than a truncation $n_c$. 
We have verified that the topological edge modes can always be observed in the regime defined by 
Eq. \ref{eq-topo2}. In the gapless topological FFLO phase regime, the gapless excitations occurs at $k \ne 0$, and
the bulk excitation gap is still finite at zero momentum. This finite energy gap ensures that the edge modes can always being directly detected 
via {\it local} momentum resolved radio frequency spectroscopy\cite{rf1, rf2, rf3}. 

We can write the edge modes near $k \rightarrow 0$ as
\begin{equation}
H_{\text{edge}} = \sum_k v_{\text{L}} \psi_{\text{L}}^\dagger k \psi_{\text{L}}^{} + v_{\text{R}} \psi_{\text{R}}^\dagger k \psi_{\text{R}}^{}
\end{equation}
where L and R correspond to the left and right edge, respectively. We find the gapless topological FFLO phase regime can be divided into two different regimes by
the sign of $v_{\text{L}} v_{\text{R}}$, where $v_{\text{L}} v_{\text{R}} > 0 $ corresponds to the co-propagating edge modes while its opposite corresponds to 
the counter-propagating edge modes, respectively. It represents the most striking feature of this new topological phase. In solid materials, $v_{\text{L}} = -v_{\text{R}}$ is always fulfilled,
thus only the counter-propagating edge modes can be observed. In some metarials with Chern number greater than one, co-propagating edge modes may be observed, however,
these modes should propagate at the same edge\cite{Galitski}. This interesting feature directly comes from the finite momentum pairing and in-plane Zeeman field.
We need to emphasize that different velocities correspond to different density of states (DOS)  near zero energy,  therefore it can be directly verified in future experiments. 

We explore the finite size effect for the same geometry defined above. The edge state splitting as a function of stripe width $L$ is presented in Fig. \ref{fig5}.  We can define
the superfluids coherence length $\xi_0 = \hbar v_{\text{F}}/E_0$,  where $v_{\text{F}}$ and $E_0$  are the Fermi velocity and energy gap at zero momentum respectively.
 In the gapped topological phase regime, we observe that the splitting decay exponentially, $\delta E \sim e^{-L/\xi}$,  where $\xi \sim \xi_0$. 
For $\Delta = 0.30$, we find $\xi = 14.2$ ($\xi_0 = 17.7$), and for $\Delta = 0.45$, $\xi = 9.0$ ($\xi_0 = 11.7$). 
In the gapless phase regime, the strong couping between different modes near $E = 0$ gives rise to oscillation of the edge state splitting, which can
be formulated as $\delta E \sim f(L)  e^{-L/\xi}$, where $f(L)$ is an nonzero oscillation function (see inset of Fig. \ref{fig5} for different truncations $n_c$). 
Such oscillation should be a typical feature of edge modes in gapless topological phase. In the trivial phase regime, no edge modes can be observed, and the gap splitting
quickly approaches the bulk gap when $L > \xi_0$ (here for $\Delta = 0.65$, $\xi_0 = 6.3$). 

To conclude, in this work we have discussed 
the basic physics in topological FFLO superfluids which still lack a direct counterpart in solid materials. The novel features of this new phase can be directly
implemented in state-of-the-art ultracold atoms in a tunable manner, including not only interaction strength, but also 
pairing asymmetry. These investigations can greatly enrich our understanding of topological matter and its basic pairing symmetry -- a fundamental issue in modern physics.

\textbf{Acknowledgement:} This work is supported by Hong Kong RGC/GRF Projects (No. 401011 and No. 2130352) and  the Chinese University of Hong Kong (CUHK) Focused Investments Scheme.


\begin{thebibliography}{99}
\bibitem{Solids} In solid materials, the Zeeman field and SOC energy are generally much smaller than the chemical potential (or Fermi energy).
\end{thebibliography}
\end{document}